\documentclass[twocolumn,,amsmath,amssymb]{revtex4}
\pdfoutput=1

\usepackage{graphicx}
\usepackage{dcolumn}
\usepackage{bm}
\newcommand{\etal}{\emph{et al.}}
\newcommand{\be}{\begin{equation}}
\newcommand{\ee}{\end{equation}}
\newcommand{\bfig}{\begin{figure}}
\newcommand{\efig}{\end{figure}}

\begin{document}      
\title{The thermal Hall effect of spin excitations in a Kagome magnet.
} 

\author{Max Hirschberger$^{1}$}
\author{Robin Chisnell$^{2}$}
\author{Young S. Lee$^2$}
\author{N. P. Ong$^{1,*}$}
\affiliation{
$^1$Department of Physics, Princeton University, Princeton, NJ 08544\\
$^2$Department of Physics, Massachusetts Institute of Technology, Cambridge, MA 02139
} 

\date{\today}      
\begin{abstract}
\end{abstract}
 
\maketitle      
{\bf At low temperatures, the thermal conductivity of spin excitations in a magnetic insulator can exceed that of phonons. However, because they are charge neutral, the spin waves are not expected to display a thermal Hall effect in a magnetic field. Recently, this semiclassical notion has been upended in quantum magnets in which the spin texture has a finite chirality~\cite{Katsura,Onose,Murakami1,Murakami2,Hirschberger}. In the Kagome lattice, the chiral term generates a Berry curvature. This results in a thermal Hall conductivity $\kappa_{xy}$ that is topological in origin. Here we report observation of a large $\kappa_{xy}$ in the Kagome magnet Cu(1-3, bdc) which orders magnetically at 1.8 K~\cite{Nocera,Keren,Chisnell}. The observed $\kappa_{xy}$ undergoes a remarkable sign-reversal with changes in temperature or magnetic field, associated with sign alternation of the Chern flux between magnon bands. We show that thermal Hall experiments probe incisively the effect of Berry curvature on heat transport~\cite{Murakami1,Murakami2,LeeHan}.    
}

\begin{figure*}[t]
\includegraphics[width=12 cm]{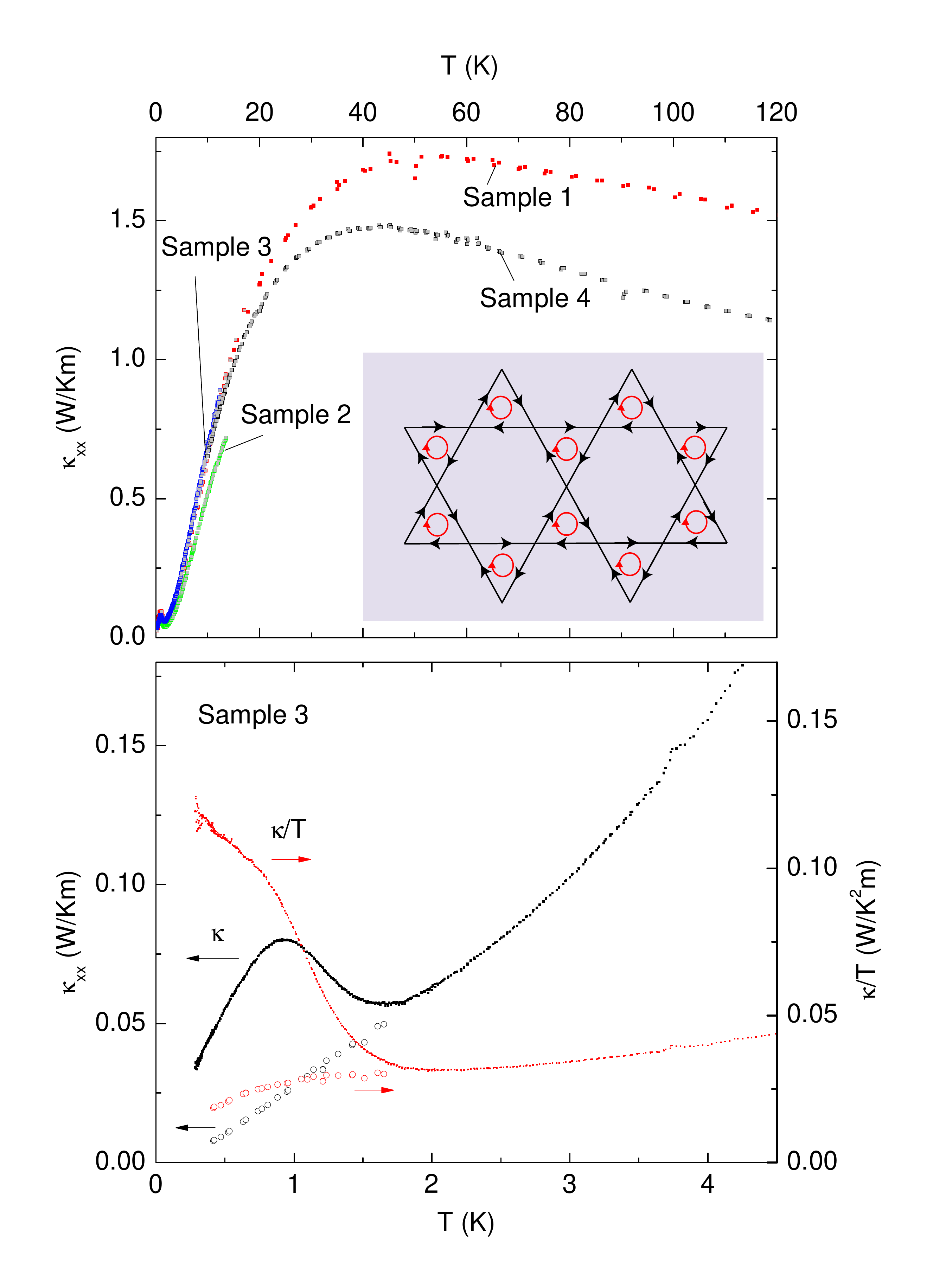}
\caption{\label{figK} 
The in-plane thermal conductivity $\kappa$ (in zero $B$) measured in the Kagome magnet Cu(1,3-benzenedicarboxylate) [Cu(1,3-bdc)]. At 40-50 K, $\kappa$ displays a broad peak followed by a steep decrease reflecting the freezing out of phonons (Panel A). The spin excitation contribution becomes apparent below 2 K. The inset is a schematic of the Kagome lattice with the LRO chiral state~\cite{Katsura}. The arrows on the bonds indicate the direction of advancing phase $\phi = \tan^{-1} D/J$. Panel B plots $\kappa$ (black symbols) and $\kappa/T$ (red) for $T<$ 4.5 K. Below the ordering temperature $T_C$ = 1.8 K, the magnon contribution to $\kappa$ appears as a prominent peak that is very $B$ dependent. Values of $\kappa$ and $\kappa/T$ at large $B$ (identified with phonons) are shown as open symbols. 
}
\end{figure*}

In magnets with strong spin-orbit interaction, competition between the Dzyaloshinskii-Moriya (DM) exchange $D$ and the Heisenberg exchange $J$ can engender canted spin textures with long-range order (LRO). Katsura, Nagaosa and Lee (KNL)~\cite{Katsura} predicted that, in the Kagome and pyrochlore lattices, the competition can lead to a state with extensive chirality $\chi = {\bf S}_i\cdot{\bf S}_j\times{\bf S}_k$ (${\bf S}_i$ is the spin at site $i$) and a large thermal Hall effect. Subsequently, Matsumoto and Murakami (MM)~\cite{Murakami1,Murakami2} amended KNL's calculation using the gravitational-potential approach~\cite{Luttinger,Streda} to relate $\kappa_{xy}$ directly to the Berry curvature. In the boson representation of the spin Hamiltonian, $\chi$ induces a complex ``hopping'' integral $t = \sqrt{J^2+D^2}\cdot e^{i\phi}$ with $\tan\phi = D/J$ (Fig. \ref{figK}A, inset)~\cite{Katsura,Murakami1,LeeHan}. Hence as they hop between sites, the bosons accumulate the phase $\phi$, which implies the existence of a vector potential $\bf A(k)$ permeating $\bf k$-space. The corresponding Berry curvature ${\bf \Omega( k) =  \nabla_{k}\times A(k)}$ acts like an effective magnetic field that imparts a transverse velocity to a magnon wave packet driven by an external gradient or force. Analogous to the intrinsic anomalous Hall effect in metals~\cite{NagaosaRMP}, the transverse velocity appears as a thermal Hall conductivity $\kappa_{xy}$ that is dissipationless and topological in origin, but now observable for neutral currents. A weak $\kappa_{xy}$ was first observed by Onose \etal~\cite{Onose} in the pyrochlore ferromagnet Lu$_2$V$_2$O$_7$. A large $\kappa_{xy}$ was recently reported in the frustrated pyrochlore Tb$_2$Ti$_2$O$_7$~\cite{Hirschberger}.

Each magnon band $n$ contributes a term to $\kappa_{xy}$ with a sign determined by the integral of $\bf \Omega( k)$ over the Brillouin zone (the Chern number). Recently, Lee, Han and Lee (LHL)~\cite{LeeHan} calculated how $\kappa_{xy}$ undergoes sign changes as the occupancy of the bands changes with $T$ or $B$. Thus, in addition to providing a powerful way to probe the Berry curvature, $\kappa_{xy}$ also yields valuable information on the magnon band occupancy. Sign changes from next-nearest neighbor exchange were studied in Ref.~\cite{Mook}.

\begin{figure*}[t]
\includegraphics[width=18 cm]{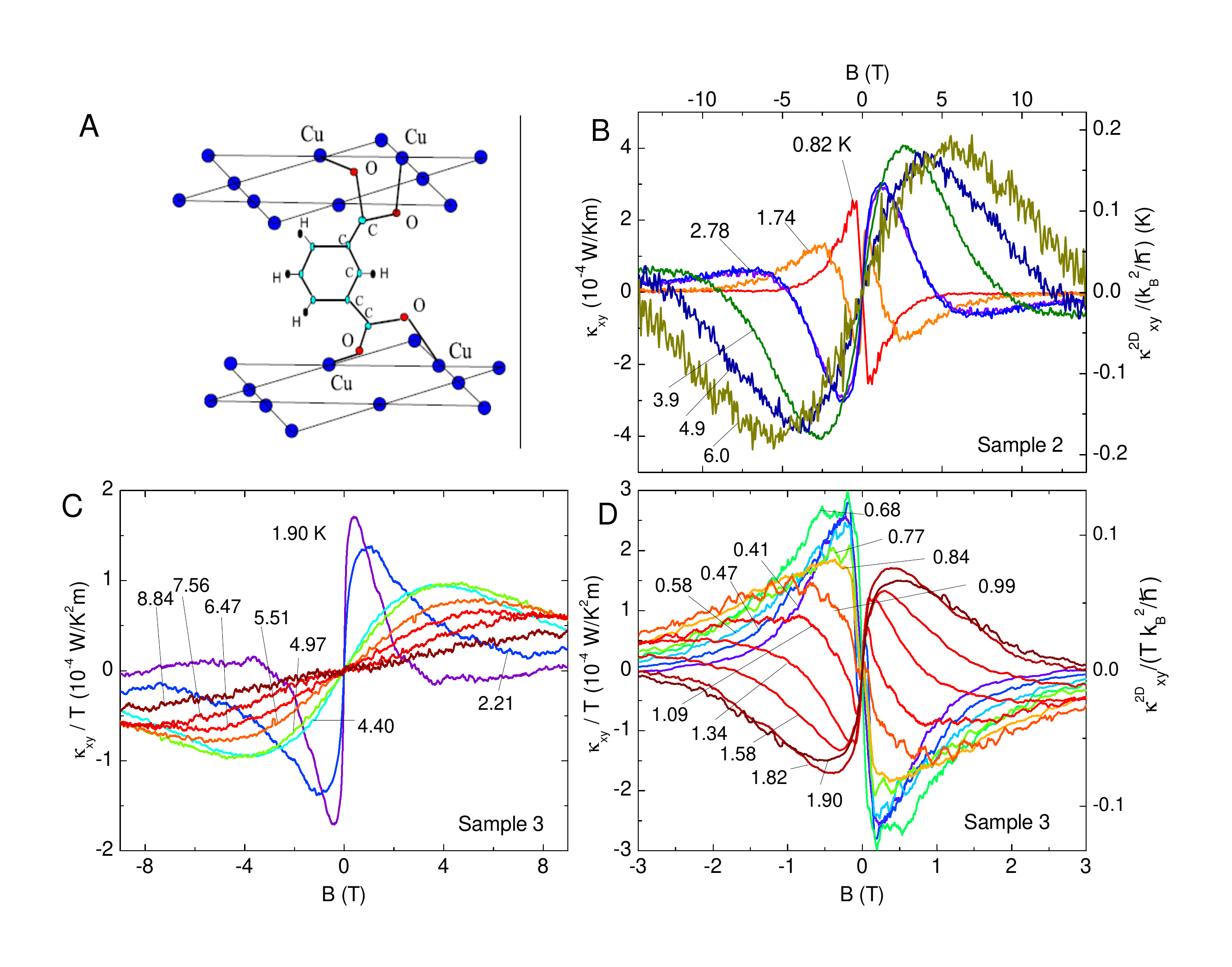}
\caption{\label{figKxy} 
The thermal Hall conductivity $\kappa_{xy}$ measured in Cu(1,3-bdc). In Panel A, the sketch shows the spin-$\frac12$ Cu ions in adjacent Kagome planes linked by a benzene molecule in which 2 of the C-H bonds have been replaced by the carboxylate ion C-C (from Ref.~\cite{Keren}). The in-plane exchange $J$ is mediated by the O-C-O bond while the weak interplane exchange $J_c$ goes through the O-5C-O bonds.  Panel B plots the strongly non-monotonic profiles of $\kappa_{xy}$ vs. $B$ in Sample 2. The dispersive profile changes sign below $\sim$ 1.7 K. The right scale gives $\kappa^{2D}/(k_B^2/\hbar)$ (per plane) obtained by multiplying $\kappa_{xy}$ by $d\hbar/k_B^2$ = 443.2 (SI units). Panel C and D show corresponding curves in Sample 3 (now plotted as $\kappa_{xy}/T$). Above $T_C$ (Panel C), $\kappa_{xy}/T$ is $p$ type. The behavior below 1.90 K is shown in Panel D. At 1.09 K, the $n$-type contribution appears in weak $B$, and eventually changes $\kappa_{xy}/T$ to $n$-type at all $B$. Right scale in D reports $\kappa_{xy}^{2D}/(Tk_B^2/\hbar)$. 
}
\end{figure*}

The Kagome magnet Cu(1,3-benzenedicarboxylate) [or Cu(1,3-bdc)] is comprised of stacked Kagome planes separated by $d$ = 7.97 $\mathrm\AA$~\cite{Nocera,Keren,Chisnell} (Fig. \ref{figKxy}A). The spin-$\frac12$ Cu$^{2+}$ moments interact via an in-plane ferromagnetic exchange $J = 0.6$ meV. Spin-orbit coupling and the absence of inversion symmetry result in a DM exchange $D = 0.09$ meV ($D/J = 0.15$), with the DM vector $\bf D$ nearly $\parallel\bf\hat{c}$~\cite{Chisnell}. The weak antiferromagnetic exchange between planes $J_c\sim 1\,\mu$eV~\cite{Chisnell}, mediated by a benzene linker via a O-5C-O bond, leads to a critical temperature $T_C\sim$ 1.8 K. In the ordered state, the magnetization $\bf M$ lies in-plane but alternates in sign between planes. A weak $B\sim$ 0.05 T is sufficient to align the moments~\cite{Chisnell}. Crystals (typically $4\times4\times 0.4\,$mm$^3$) grow as transparent, bright blue, hexagonal platelets with the largest face parallel to the Kagome planes (spanned by vectors $\bf a$ and $\bf b$). The thermal conductivity tensor $\kappa_{ij}$ is defined by $J^Q_i = \kappa_{ij}(-\partial_j T)$, with ${\bf{J}}^Q$ the thermal current density. We apply  the thermal gradient $-\nabla T \parallel \hat{\textbf{x}}\parallel(\bf a, b)$, with $\bf B\parallel \hat{\textbf{z}}\parallel {\textbf{c}}$.

As we cool the sample in zero $B$, the thermal conductivity $\kappa$ (nearly entirely from phonons) initially rises to a very broad peak at 45 K (Fig. \ref{figK}A). Below the peak, $\kappa$ decreases rapidly as the phonons freeze out. Starting near 10 K, the spin contribution $\kappa^s$ becomes apparent. As shown in Fig. \ref{figK}B, this leads to a minimum in $\kappa$ near $T_C$ followed by a large peak at $\sim\frac12 T_C$. Factoring out the entropy, we find that $\kappa/T$ (red curve) increases rapidly below $T_C$. This reflects the increased stiffening of the magnon bands as LRO is established. Below 800 mK, the increase in $\kappa/T$ slows to approach saturation. The open black circles represent the phonon conductivity $\kappa_{ph}$ deduced from the large-$B$ values of $\kappa_{xx}(T,H)$ (see below). Likewise, $\kappa_{ph}/T$ is plotted as open red circles. The difference $\kappa - \kappa_{ph}$ is the estimated thermal conductivity of magnons $\kappa^s$ in zero $B$.

Given that Cu(1,3-bdc) is a transparent insulator, it exhibits a surprisingly large thermal Hall conductivity (Fig. \ref{figKxy}). Above $T_C$, the field profile of $\kappa_{xy}$ is non-monotonic, showing a positive peak at low $B$, followed by a zero-crossing at higher $B$ (see curve at 2.78 K in Fig. \ref{figKxy}B). We refer to a positive $\kappa_{xy}$ as ``$p$-type''. Below $T_C$, an interesting change of sign is observed (curves at 1.74 and 0.82 K). The weak hysteresis, implying a coercitive field $<$1500 Oe at the lowest temperatures, is discussed in the supplementary information (SI). This sign-change is investigated in greater detail in Sample 3 (we plot $\kappa_{xy}/T$ in Fig. \ref{figKxy}C). The curves of $\kappa_{xy}/T$ above $T_C$ are similar to those in Sample 2. As we cool towards $T_C$, the peak field $H_p$ decreases rapidly, but remains resolvable below $T_C$ down to 1 K (Fig. \ref{figKxy}D). However, below 1 K, the $p$-type curve is swamped by a rising $n$-type contribution that eventually dominates below 0.6 K. Some features are in qualitative agreement with the calculations of LHL~\cite{LeeHan}, but there are significant differences as well.

\begin{figure*}[t]
\includegraphics[width=18 cm]{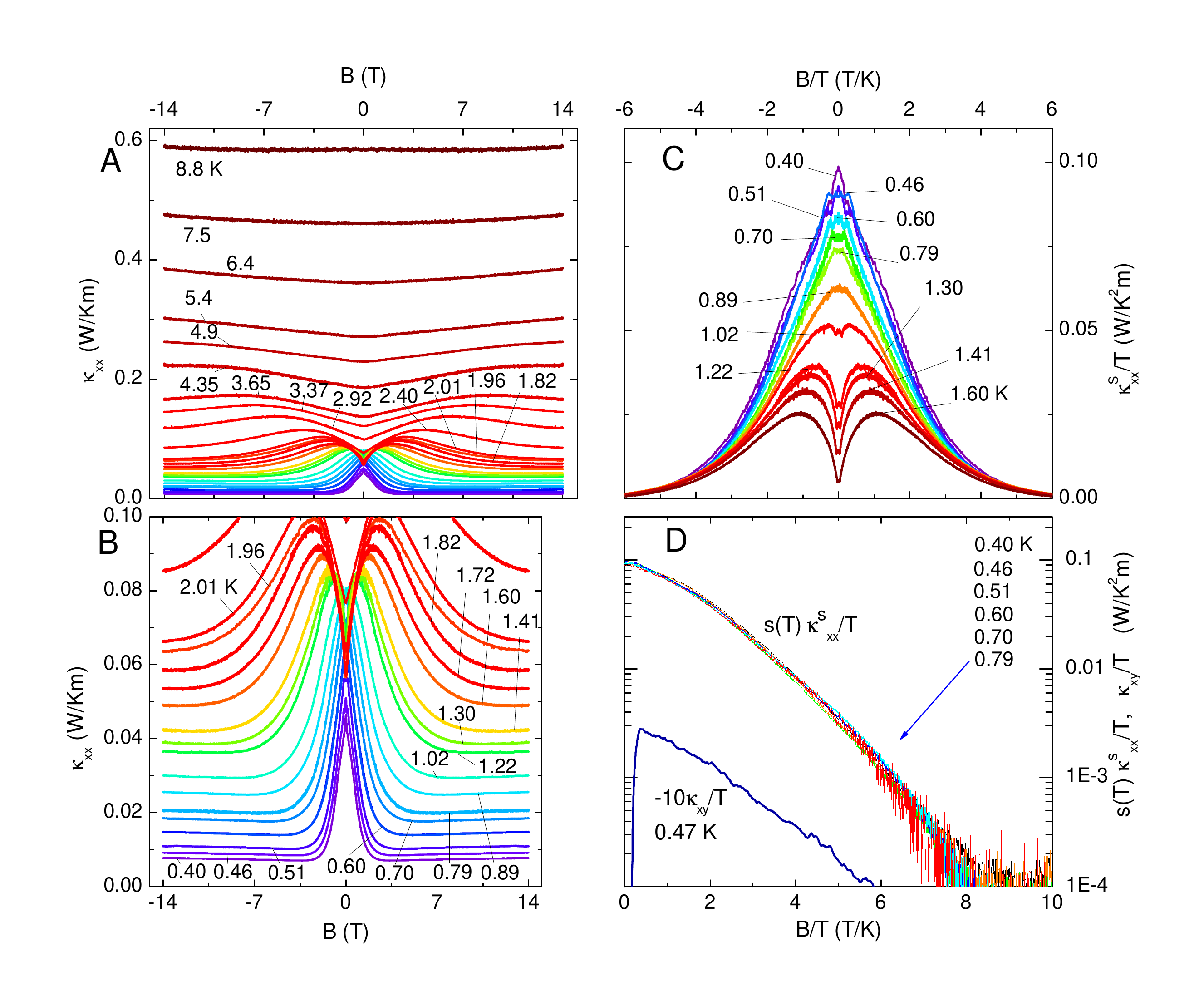}
\caption{\label{figKxx} 
The effect of field $B$ on $\kappa_{xx}$ and scaling behavior at low $T$, for sample 3. The curves in Panel A show that the $B$-dependence of $\kappa_{xx}$ is resolved (in the range $|B|<$ 14 T) only at $T<\sim$6.5 K. The expanded scale in Panel B shows that, near $T_C$ (1.8 K), $\kappa_{xx}$ has a non-monotonic profile with a $V$-shaped minimum at $B$ = 0 (identified with stiffening of the magnon bands by the field). Below 1 K, however, $\kappa_{xx}$ has a strictly monotonic profile that terminates in a sharp cusp peak as $B\to$0. At each $T<T_C$, the constant ``floor'' profile at large $B$ is identified with $\kappa_{ph}$. Panel C shows that the complicated pattern in Panel B simplifies when plotted as $\kappa^s_{xx}/T$ vs. $B/T$. Further, multiplying by a scaling factor $s(T)$ collapses all the curves below 1 K to a ``universal'' curve, as shown on log scale in Panel D. The slope at large $B$ gives a Zeeman gap with $g$ = 1.6. The Hall curve $-\kappa_{xy}/T$ has a similar slope at large $B$.  
}
\end{figure*}

To relate the thermal Hall results to magnons, we next examine the effect of $B$ on the longitudinal thermal conductivity $\kappa_{xx}$. As shown in Fig. \ref{figKxx}A, $\kappa_{xx}$ is initially $B$-independent for $T>$ 10 K, suggesting negligible interaction between phonons and the spins. The increasingly strong $B$ dependence observed below 4 K is highlighted in Fig. \ref{figKxx}B. Despite the complicated evolution of the profiles, all the curves share the feature that the $B$-dependent part is exponentially suppressed at large $B$, leaving a $B$-independent ``floor'' which we identify with $\kappa_{ph}(T)$ (plotted as open symbols in Fig. \ref{figK}B). Subtracting the floor allows the thermal conductivity due to spins to be defined as
$\kappa^s_{xx}(T,H) \equiv \kappa_{xy}(T,H) - \kappa_{ph}(T).$
The exponential suppression in all curves is more apparent if we replot the spin part as $\kappa^s_{xx}/T$ vs. $B/T$ (Fig. \ref{figKxx}C). As may be seen, the aymptotic form at large $B$ in all curves depends only on $B/T$. 

In the interval 0.9 K$\to T_C$, $\kappa^s_{xx}$ displays a $V$-shaped minimum at $B$ = 0 followed by a peak at the field $H_p(T)$. Since $\kappa^s$ (at $B=0$) is falling rapidly within this interval due to softening of the magnon bands (see Fig. \ref{figK}B), we associate the $V$-shaped profile with stiffening of the magnon bands by the applied $B$. At low enough $T$ ($<$0.8 K), this stiffening is unimportant and the curves are strictly monotonic. We find that they follow the same universal form. To show this, we multiply each curve by a $T$-dependent scale factor $s(T)$ and plot them on semilog scale in Fig. \ref{figKxx}D. 
In the limit of large-$B$, the universal curve follows the activated form 
\be
\kappa^s_{xx}\to T e^{-\beta \Delta}, 
\label{eq:ks}
\ee
with the Zeeman gap $\Delta = g\mu_B B$ where $\beta = 1/k_BT$, $\mu_B$ is the Bohr magneton, and $g$ the g-factor. The inferred value of $g$ ($\sim$1.6) is consistent with the Zeeman gap measured in a recent neutron scattering experiment. 

For comparison, we have also plotted $-\kappa_{xy}/T$ (at 0.47 K) in Fig. \ref{figKxx}D. Within the uncertainty, it also decreases exponentially at large $B$ with a slope close to $\Delta$. Hence the exponential suppression of the magnon population resulting from $\Delta$ is evident in both $\kappa^s_{xx}$ and $\kappa_{xy}$. This provides strong evidence that the observed thermal Hall signal is intrinsic to the magnons.

According to MM~\cite{Murakami1,Murakami2}, $\kappa_{xy}$ (in 2D) is given by
\be
\kappa_{xy} = \frac{2k_B^2 T}{\hbar V}\sum_{n,{\bf k}} \; c_2(\rho_n)\Omega_n({\bf k}) 
\label{kxy}
\ee
where $c_2(x) = (1+x)(\log \frac{1+x}{x})^2 - (\log x)^2 - 2\rm{Li}_2(-x)$, with $\rm{Li}_2(x)$ the polylogarithmic function, and $\rho_n = 1/[e^{\beta(E_n-\mu)}-1]$ is the Bose-Einstein distribution, with $E_n$ the magnon energy in band $n$ and $\mu$ the chemical potential.

LHL~\cite{LeeHan} have calculated $\kappa_{xy}(T,B)$ applying the Holstein-Primakoff (HP) representation below and above $T_C$, and Schwinger bosons (SB) above $T_C$. In the ordered phase, the HP curves capture the sign changes observed in $\kappa_{xy}(T,H)$: a purely $n$-type curve at the lowest $T$ and, closer to $T_C$, a sign-change induced by a $p$-type term. Moreover, the calculated curves at each $T$ exhibit the high-field suppression, in agreement with Fig. \ref{figKxx}D. For Sample 3, the peak values of $\kappa_{xy}^{2D}$ are in agreement with the HP curves (0.04 K at $T$ = 0.4 K; 0.2 K at 4.4 K). In the paramagnetic region, however, our field profiles disagree with the SB curves. Above $T_C$, $\kappa_{xy}$ is observed to be $p$-type at all $B$ whereas the SB curves are largely $n$-type apart from a small window at low $B$. The comparison suggests that the HP approach is a better predictor than the SB representation even above $T_C$.

In addition to confirming the existence of a large $\kappa_{xy}$ in the Kagome magnet (and hence a finite, extensive $\chi$), the measured $\kappa_{xy}$ can be meaningfully compared with detailed calculations based on the Berry curvature. For chiral magnets, $\kappa_{xy}$ is capable of probing incisively the effect of the Berry curvature on transport currents.


\vspace{1cm}\noindent
{\bf Acknowledgements}\\
We acknowledge support from the US National Science Foundation through the MRSEC grant DMR 1420541. N.P.O. was supported by US Army Office of Research (contract W911NF-11-1-0379) and by the Gordon and Betty Moore Foundation’s EPiQS Initiative through Grant GBMF4539.

\newpage

\begin{thebibliography}{99}

\bibitem{Katsura} H. Katsura, N. Nagaosa and P. A. Lee,
``Theory of the Thermal Hall Effect in Quantum Magnets,"
Phys Rev Lett {\bf 104}, 066403 (2010).


\bibitem{Onose} Y. Onose, T. Ideue, H. Katsura, Y. Shiomi, N. Nagaosa and Y. Tokura,
``Observation of the Magnon Hall Effect,'' 
Science {\bf 329}, 297-299 (2010).

\bibitem{Murakami1} R. Matsumoto and S. Murakami,
``Theoretical Prediction of a Rotating Magnon Wave Packet in Ferromagnets,'' 
Phys Rev Lett {\bf 106}, 197202 (2011).

\bibitem{Murakami2} R. Matsumoto and S. Murakami,
``Rotational motion of magnons and the thermal Hall effect,''
Phys Rev B {\bf 84}, 184406 (2011).

\bibitem{Hirschberger} Max Hirschberger, Jason W. Krizan, R. J. Cava and N. P. Ong, 
``Large thermal Hall conductivity of neutral spin excitations in a frustrated quantum magnet,''
\emph{submitted}.

\bibitem{Nocera} Emily A. Nytko, Joel S. Helton, Peter M\"{u}ller, and Daniel G. Nocera,
``A Structurally Perfect S = $\frac12$ Metal-Organic Hybrid Kagome Antiferromagnet,''
J. Am. Chem. Soc. {\bf 130}, 2922-2923 (2008).

\bibitem{Keren} Lital Marcipar, Oren Ofer, Amit Keren, Emily A. Nytko, Daniel G. Nocera, Young S. Lee, Joel S. Helton, and Chris Bains,
``Muon-spin spectroscopy of the orgnaometallic spin-$\frac12$ kagome-lattice compound Cu(1,3-benzenedicarboxylate),''
Phys. Rev. B {\bf 80}, 132402 (2009).

\bibitem{Chisnell} Robin Chisnell, PhD. thesis, MIT



\bibitem{LeeHan} Hyunyong Lee, Jung Hoon Han and Patrick A. Lee,
``Thermal Hall Effect of Spins in a Paramagnet,''
arXiv: 1410.3759v2.

\bibitem{Mook} Alexander Mook, J\"{u}rgen Henk and Ingrid Mertig,
``Magnon Hall effect and topology in kagome lattices: A theoretical investigation,''
Phys. Rev. B {\bf 89}, 134409 (2014). 


\bibitem{NagaosaRMP} Naoto Nagaosa, Jairo Sinova, Shigeki Onoda, A. H. MacDonald, and  N. P. Ong, 
``Anomalous Hall Effect,'' 
Rev. Mod. Phys. {\bf 82}, 1539 (2010).


\bibitem{Luttinger} J. M. Luttinger,
``Theory of Thermal Transport Coefficients,''
Phys. Rev. {\bf 135}, A1505-A1514 (1964).

\bibitem{Streda} H. Oji and P. Streda,
``Theory of Electronic Thermal Transport - Magnetoquantum Corrections to the Thermal Transport-Coefficients,''
Phys Rev B {\bf 31}, 7291-7295 (1985).



\end{thebibliography}
\end{document}